\documentclass[12pt]{article}
\usepackage{a4}
\usepackage{amsmath}
\usepackage{amssymb}
\usepackage{amsfonts}
\usepackage{graphics}
\usepackage{epsfig}
\usepackage{latexsym}
\usepackage{eepic}

\begin{document}

\noindent hep-th/
   \hfill  November 2010 \\

\renewcommand{\theequation}{\arabic{section}.\arabic{equation}}
\thispagestyle{empty}
\vspace*{-1,5cm}
\noindent \vskip3.3cm

\begin{center}
{\Large\bf On Nonlinear Higher Spin Curvature}

{\large Ruben Manvelyan${}^{ab}$\footnote{manvel@physik.uni-kl.de}, Karapet Mkrtchyan${}^{ab}$\footnote{karapet@yerphi.am} ,\\Werner R\"uhl${}^{a}$\footnote{ruehl@physik.uni-kl.de} and Murad Tovmasyan${}^{b}$}\footnote{mtovmasyan@ysu.am}

\medskip

${}^{a}${\small\it Department of Physics\\ Erwin Schr\"odinger Stra\ss e \\
Technical University of Kaiserslautern, Postfach 3049}\\
{\small\it 67653
Kaiserslautern, Germany}\\
\medskip
${}^{b}${\small\it Yerevan Physics Institute\\ Alikhanian Br.
Str.
2, 0036 Yerevan, Armenia}\\
\medskip
\end{center}\vspace{2cm}

\bigskip
\begin{center}
{\sc Abstract}
\end{center}
\quad
We present the first nonlinear term of the higher spin curvature which is covariant with respect to deformed gauge transformations that are linear in the field. We consider the case of spin 3 after presenting spin 2 as an example, and then construct the general spin $s$ quadratic term of the deWit-Freedman curvature.

\newpage

\section{Introduction}

\quad Geometry plays a very important role in field theory. General relativity is the most beautiful and obvious example to prove this. In higher spin gauge field theory the elements of geometry were studied first in the classical paper by Bernard de Witt and Daniel Freedman \cite{DF}.
Geometrical interpretations of Yang-Mills theory and general relativity hint on possible geometric structures of higher spin gauge field theories in general \cite{DF}, \cite{Francia:2002aa}, \cite{Damour}, \cite{Francia}, \cite{VasilievEqn}. For the well analyzed existing theories gauge symmetry itself has a hidden geometrical origin. Since the gauge symmetry is a redundancy in the (non-observable) variables of the physical system, the observables of any gauge theory are gauge invariant. For electrodynamics the observables are the components of the Maxwell tensor $F_{\mu\nu}$ (curvature of the electromagnetic field $A_\mu$), for gravity these are the components of the Riemann curvature tensor $R_{\mu\nu\lambda\rho}$ (constructed from the dynamical field, the metric tensor $g_{\mu\nu}$). In analogy with these cases, in the higher spin gauge field theory the curvature of the higher spin field can be connected to observables. The linear curvature as well as linearized ``Cristoffel symbols'' and connections for any higher spin field are known from \cite{DF}. Taking traces of the linearized curvature one can get geometrical free equations of motion for the unconstrained higher spin gauge field \cite{Francia:2002aa}, \cite{Bekaert:2003az}, which is nonlocal, but can be localized with the help of partial gauge fixing after which the geometrical equation coincides with Fronsdal's equation \cite{Frons}.

There are two most common covariant approaches to the task under consideration. The first one is the frame-like formalism, which is developed by Vasiliev and coauthors (see \cite{Bekaert:2005vh} for a review). The second approach is a generalization of linearized gravity - the metric-like theory of higher spin fields \cite{Frons} (for recent development see \cite{Manvelyan:2010je} and references therein). We use the second one.

In this letter we construct a higher spin curvature that is of second order in the field, which can be used to find that part of the geometric equation of motion that is quadratic in the field, which, after the same partial gauge fixing (ruling out the trace of the gauge parameter and the double trace of the field), may or may not coincide with the corresponding second order of Fronsdal's equation of motion corresponding to the cubic Lagrangian derived in \cite{Manvelyan:2010jr}. The cubic (self)interaction Lagrangian and the corresponding gauge transformation laws for Fronsdal's fields that are of first order on field, are known \cite{Manvelyan:2010jr} - \cite{Mkrtchyan:2010pp}. The connection between these two independent constructions is a subject for future investigations. Now we are going to construct the first nonlinear term in the higher spin curvature independently, and with no use of Fronsdal's constraints on the fields involved. In order to get a gauge covariant curvature, we don't need to impose the constraint of tracelessness on the gauge parameter respectively the constraint of double tracelessness on the higher spin field. We don't make use of Fronsdal's theory in this letter, because the nonlinear curvature that we construct here, describes not only Fronsdal's (constrained) fields, but, if it exists to all orders in the field, gives us a possibility for the construction of theories of Higher Spin fields alternative to Fronsdal's (consider the nonlinear continuation of conformal higher spin theory \cite{Fradkin:1985am, Segal:2002gd} as an example).

In Section 2 we discuss as a simple example linearized gravity and construct the second order of the Riemann curvature using the gauge covariance condition. In Section 3 we construct the curvature of second order in the field for the spin 3 gauge field. In the final Section 4 we derive that part of the nonlinear gauge covariant curvature that is of second order in the field for any higher spin field which reproduces all lower spin cases and has the same symmetry properties as the linear curvature.

The most remarkable property of the curvature that is of second order in the field is that it is a sum of terms, which are all squares of the linear connection $\Gamma^{(s-1)}$, the s minus first member in the hierarchy of de Witt - Freedman - Christoffel symbols \cite{DF}. Using shorthands explained later, we can summarize the curvature up to second order in the field in the following compact way
\begin{equation}
R=d\Gamma+\Gamma \star \Gamma \label{1.1}
\end{equation}
for any higher spin gauge field.
It is worth noting at the end of this introduction that our result for the second order curvature involves only the square of generalized Christoffel symbols constructed from the same higher spin gauge field. In other words, on this stage of the curvature construction there is no contribution coming from any other higher spin field which could be found in the frame-like formulation of \cite{VasilievEqn}. The explanation of this phenomena is the following: Quadratic deformations of the de Witt - Freedman curvature correspond to the cubic self-interaction of the higher spin gauge fields and the linear deformation of the gauge algebra \cite{vanDam1, Manvelyan:2010wp}. On the linear level the closure of the gauge transformation algebra ($[\delta_{(0)},\delta_{(1)}]\sim \delta_{(0)}$) needs no share of other higher spin field transformations (see \cite{vanDam1} for the original discussion and \cite{Manvelyan:2010wp} for additional explanations). As a result the cubic selfinteraction can always be formulated as a local interaction for each separate spin \cite{Manvelyan:2010wp}, and our second order curvature is a realization of the same level of the gauge symmetry algebra. This nice property ceases to hold on the next stage of curvature or selfinteraction because the commutator of two first order gauge transformations ($[\delta_{(0)},\delta_{(2)}]+[\delta_{(1)},\delta_{(1)}]\sim \delta_{(1)}$) doesn't close without additional contributions of all other higher spin fields \cite{vanDam1, Bekaert:2010hp}. What can happen with the next order curvature and how is it connected with the corresponding frame like formulation involving other spin contributions \cite{VasilievEqn} we leave for future investigations\footnote{ Discussion of a possible mixture of different spins in the next order of curvature as well as possible continuation to the higher orders of interaction and generalization to AdS background should be studied in the future. These tasks are closely related.}.

\section{Quadratic term of linearized Riemann curvature}
\setcounter{equation}{0}
\quad We start from a consideration of the linearized Riemann curvature (spin 2 case). We use the following self consistency definitions for covariant derivatives, Christoffel symbols and curvature:
\begin{eqnarray}
  && \nabla_{\mu}V_{\nu}=\partial_{\mu}V_{\nu}+ \Gamma^{\rho}_{\mu\nu}V_{\rho} \label{2.1}\\
  && [\nabla_{\mu}, \nabla_{\nu}]V_{\lambda}=R^{\quad\,\,\,\,\rho}_{\mu\nu,\lambda}V_{\rho}\label{2.2}\\
  && \Gamma_{\rho,\mu\nu}=g_{\rho\sigma}\Gamma^{\sigma}_{\mu\nu}=\frac{1}{2}(\partial_{\rho}g_{\mu\nu}- \partial_{\mu}g_{\nu\rho}-\partial_{\nu}g_{\mu\rho})\label{2.3}\\
  && R^{\quad\,\,\,\,\rho}_{\mu\nu,\lambda}=R_{\mu\nu,\lambda\delta}g^{\delta\rho}=\partial_{\mu}\Gamma^{\rho}_{\nu\lambda}
  -\partial_{\nu}\Gamma^{\rho}_{\mu\lambda}
  +\Gamma^{\sigma}_{\mu\lambda}\Gamma^{\rho}_{\nu\sigma}-\Gamma^{\sigma}_{\nu\lambda}\Gamma^{\rho}_{\mu\sigma}\label{2.4}
\end{eqnarray}
Then we note that expression (\ref{2.3}) is very convenient for a linearization and we can obtain a linearized Christoffel symbol just replacing the metric $g_{\mu\nu}$ by the linearized field $h_{\mu\nu}=g_{\mu\nu}-\eta_{\mu\nu}$, where $\eta_{\mu\nu}$ is the Minkowski metric. To obtain the corresponding  expression for the curvature ready for linearization we can use (\ref{2.3}) and after some algebra write the full covariant curvature $R_{\mu\nu,\lambda\rho}$ in the form
\begin{eqnarray}
  && R_{\mu\nu,\lambda\rho}=\partial_{\mu}\Gamma_{\rho,\nu\lambda}- \partial_{\nu}\Gamma_{\rho,\mu\lambda}-g^{\sigma\delta}\left(\Gamma_{\sigma,\mu\lambda}\Gamma_{\delta,\nu\rho}
  -\Gamma_{\sigma,\nu\lambda}\Gamma_{\delta,\mu\rho}\right)\label{2.5}
\end{eqnarray}
Then substituting in (\ref{2.5})
\begin{eqnarray}
  && g^{\mu\nu}=\eta^{\mu\nu}-h^{\mu\nu}+h^{\mu}_{\sigma}h^{\nu\sigma}- \dots \label{2.6}\\
  &&  2\Gamma_{\rho,\mu\nu}=\partial_{\rho}h_{\mu\nu}- \partial_{\mu}h_{\nu\rho}-\partial_{\nu}h_{\mu\rho}\label{2.7}
\end{eqnarray}
we arrive at the following expansion of the curvature up to the third order in the field:
\begin{eqnarray}
  && R^{h}_{\mu\nu,\lambda\rho}= R^{(1)}_{\mu\nu,\lambda\rho}+R^{(2)}_{\mu\nu,\lambda\rho}+R^{(3)}_{\mu\nu,\lambda\rho} + \dots \label{2.8}\\
  && 2 R^{(1)}_{\mu\nu,\lambda\rho}=\partial_{\mu}\partial_{\rho}h_{\nu\lambda}- \partial_{\nu}\partial_{\rho}h_{\mu\lambda}-\partial_{\mu}\partial_{\lambda}h_{\nu\rho}
  +\partial_{\nu}\partial_{\lambda}h_{\mu\rho} \label{2.9}\\
  && R^{(2)}_{\mu\nu,\lambda\rho}=-\eta^{\sigma\delta}\left(\Gamma_{\sigma,\mu\lambda}\Gamma_{\delta,\nu\rho}
  -\Gamma_{\sigma,\nu\lambda}\Gamma_{\delta,\mu\rho}\right) \label{2.10}\\
  &&R^{(3)}_{\mu\nu,\lambda\rho} = h^{\sigma\delta}\left(\Gamma_{\sigma,\mu\lambda}\Gamma_{\delta,\nu\rho}
  -\Gamma_{\sigma,\nu\lambda}\Gamma_{\delta,\mu\rho}\right)\label{2.11}
\end{eqnarray}

Finalizing this section we note that the same expansion could be recovered from the initial linearized curvature (\ref{2.9}) and gauge invariance of order zero in the field $\delta_{(0)}h_{\mu\nu}=\partial_{\mu}\epsilon_{\nu}+\partial_{\nu}\epsilon_{\mu}$ using the "covariant" (not invariant) Noether's equation:
    \begin{equation}\label{2.12}
        \delta_{(1)}R^{(1)}_{\mu\nu,\lambda\rho}+\delta_{(0)}R^{(2)}_{\mu\nu,\lambda\rho}=
        \mathcal{L}_{\epsilon}R^{(1)}_{\mu\nu,\lambda\rho}
    \end{equation}
where $\mathcal{L}_{\epsilon}R^{(1)}_{\mu\nu,\lambda\rho}$ is the Lie derivative of the first order curvature
\begin{eqnarray}
\mathcal{L}_{\epsilon}R^{(1)}_{\mu\nu,\lambda\rho} = &&\varepsilon
^\rho  \partial _\rho R_{\alpha \beta ,\mu \nu }^{(1)}  + \partial
_\alpha \varepsilon ^\rho  R_{\rho \beta ,\mu \nu }^{(1)}
+
\partial _\beta \varepsilon ^\rho  R_{\alpha \rho ,\mu \nu }^{(1)}
\nonumber\\&&+
\partial _\mu \varepsilon ^\rho  R_{\alpha \beta ,\rho \nu }^{(1)}
+ \partial _\nu \varepsilon ^\rho  R_{\alpha \beta ,\mu \rho }^{(1)}\label{2.13}
\end{eqnarray}
It is easy to see that as a solution of the equation (\ref{2.12}) we will find (\ref{2.10}) and the following gauge transformation
of first order in the field:
\begin{equation}\label{2.14}
    \delta_{(1)}h_{\mu\nu}=\epsilon^{\lambda}\Gamma_{\lambda,\mu\nu}+ \partial_{\mu}f_{\nu}(h,\epsilon)+\partial_{\nu}f_{\mu}(h,\epsilon)
\end{equation}
where $f_{\mu}(h,\epsilon)$  is at this stage an arbitrary vector function linear in the field $h$ and the parameter $\epsilon$.
The linearized curvature (\ref{2.9}) is invariant with respect to any gradient transformation $\delta h_{\mu\nu}=\partial_{\mu}f_{\nu}+\partial_{\nu}f_{\mu}$ with the arbitrary vector parameter $f_{\mu}$. Nevertheless the next order of Noether's procedure fixes this ambiguity, and we obtain the well known gravitational gauge transformation $\delta_{(1)} h_{\mu\nu}=\epsilon^{\rho}\partial_{\rho}h_{\mu\nu}
+\partial_{\mu}\epsilon^{\rho}h_{\nu\rho}+\partial_{\nu}\epsilon^{\rho}h_{\mu\rho}$ ($f_{\mu}=h_{\mu\rho}\epsilon^{\rho}$) in the first order on the field.

In the next sections we generalize this covariant Noether's procedure for the spin $3$ and the general spin $s$ case constructing the unknown quadratic part of the higher spin curvature.

\section{The Case of Spin 3}
\setcounter{equation}{0}
\quad To handle the spin 3 case we should introduce an additional nonabelian charge to avoid trivialization of the theory \cite{vanDam1}. Then our spin three field $h^{a}_{\alpha\beta\gamma}$ and the symmetry parameter $\varepsilon _{\beta \gamma}^a$ carry
an additional Lie algebra basis index. Introducing the corresponding zero order gauge transformation in the field
\begin{equation}\label{3.1}
\delta _\varepsilon ^0  h _{\alpha \beta \gamma }^a  =
3\partial _{(\alpha } \varepsilon _{\beta \gamma )}^a  = \partial
_\alpha  \varepsilon _{\beta \gamma }^a  + \partial _\beta
\varepsilon _{\alpha \gamma }^a  + \partial _\gamma  \varepsilon
_{\alpha \beta }^a
\end{equation}
we define the first order curvature
\begin{eqnarray}
R_{\alpha \alpha ',\beta \beta ',\gamma \gamma '}^{a(1)}=&&\partial
_\alpha  \partial _\beta  \partial _\gamma   h _{\alpha '\beta
'\gamma '}^a  \nonumber\\
&&-\partial _{\alpha '} \partial _\beta
\partial _\gamma   h _{\alpha \beta '\gamma '}^a  - \partial
_\alpha\partial _{\beta '} \partial _\gamma   h _{\alpha '\beta
\gamma '}^a  -\partial _\alpha\partial _\beta\partial _{\gamma'}
 h _{\alpha '\beta '\gamma }^a\nonumber\\
 &&+ \partial _{\alpha'} \partial _{\beta '} \partial _\gamma   h _{\alpha \beta
\gamma '}^a  +
\partial _{\alpha '} \partial _\beta  \partial _{\gamma '}  h
_{\alpha \beta '\gamma }^a  + \partial _\alpha  \partial _{\beta '}
\partial _{\gamma '}  h _{\alpha '\beta \gamma }^a\nonumber\\
&&- \partial _{\alpha '} \partial _{\beta '} \partial _{\gamma '}
 h _{\alpha \beta \gamma }^a \label{3.2}
\end{eqnarray}
from the standard condition of gauge invariance
\begin{equation}
\delta^{0}_\varepsilon
R^{a(1)}_{\alpha\alpha',\beta\beta',\gamma\gamma'}=0\label{3.3}
\end{equation}
Turning to the next step we should solve the following Noether's equation
\begin{equation}\label{3.4}
    \delta^{0}_\varepsilon
R^{a(2)}_{\alpha\alpha',\beta\beta',\gamma\gamma'}+\delta^{1}_\varepsilon
R^{a(1)}_{\alpha\alpha',\beta\beta',\gamma\gamma'}=0 +O(R^{a(1)}_{\alpha\alpha',\beta\beta',\gamma\gamma'},\varepsilon ^{b\mu \nu })
\end{equation}
where $O(R^{a(1)}_{\alpha\alpha',\beta\beta',\gamma\gamma'},\varepsilon ^{b\mu \nu })$ represents some expression that is linear in the first order curvature and the gauge parameter and has two derivatives, which plays the role of the spin 3 generalization of the Lie derivative in (\ref{2.12}).

To present a solution of (\ref{3.4}) we should first introduce the spin 3 generalization of the Christoffel symbol  (\ref{2.7}):
\begin{eqnarray}
\Gamma^{a}_{\mu\nu,\alpha\beta\gamma}&&=\partial_\mu\partial_\nu h_{\alpha\beta\gamma}^a\nonumber\\
&&-\frac{1}{2}\partial_\alpha\partial_\mu h^{a}_{\nu\beta\gamma}-\frac{1}{2}\partial_\alpha\partial_\nu h^{a}_{\mu\beta\gamma}
-\frac{1}{2}\partial_\beta\partial_\mu h^{a}_{\alpha\nu\gamma}\nonumber\\
&&-\frac{1}{2}\partial_\beta\partial_\nu h^{a}_{\alpha\mu\gamma}-\frac{1}{2}\partial_\gamma\partial_\mu h^{a}_{\alpha\beta\nu}
-\frac{1}{2}\partial_\gamma\partial_\nu h^{a}_{\alpha\beta\mu}\nonumber\\
&&+\partial_\alpha\partial_\beta h^a_{\mu\nu\gamma}+\partial_\alpha\partial_\gamma h^a_{\mu\beta\nu}+\partial_\beta\partial_\gamma h^a_{\alpha\mu\nu}\label{3.5}
\end{eqnarray}
This expression differs from the second generalized Christoffel symbol of deWit-Freedman \cite{DF} only by an additional Lie algebra index.
Then we can present the following expressions for the second order curvature
\begin{eqnarray}
R^{a(2)}_{\alpha\alpha',\beta\beta',\gamma\gamma'}&&=
f^{abc}(\Gamma^{b\mu\nu}_{\alpha\beta\gamma}\Gamma^{c}_{\mu\nu,\alpha'\beta'\gamma'}\nonumber\\
&&-\Gamma^{b\mu\nu}_{\alpha'\beta\gamma}\Gamma^{c}_{\mu\nu,\alpha\beta'\gamma'}
-\Gamma^{b\mu\nu}_{\alpha\beta'\gamma}\Gamma^{c}_{\mu\nu,\alpha'\beta\gamma'}
-\Gamma^{b\mu\nu}_{\alpha\beta\gamma'}\Gamma^{c}_{\mu\nu,\alpha'\beta'\gamma}\nonumber\\
&&+\Gamma^{b\mu\nu}_{\alpha'\beta'\gamma}\Gamma^{c}_{\mu\nu,\alpha\beta\gamma'}
+\Gamma^{b\mu\nu}_{\alpha'\beta\gamma'}\Gamma^{c}_{\mu\nu,\alpha\beta'\gamma}
+\Gamma^{b\mu\nu}_{\alpha\beta'\gamma'}\Gamma^{c}_{\mu\nu,\alpha'\beta\gamma}\nonumber\\
&&-\Gamma^{b\mu\nu}_{\alpha'\beta'\gamma'}\Gamma^{c}_{\mu\nu,\alpha\beta\gamma})\label{3.6}
\end{eqnarray}
and the first order gauge transformation
\begin{eqnarray}
\delta _\varepsilon ^1  h _{\alpha \beta \gamma }^a=&&f^{abc}
(\varepsilon ^{b\mu \nu } \partial _\mu  \partial _\nu  h
_{\alpha \beta \gamma }^c  \nonumber\\&&+ \partial _\alpha
\varepsilon ^{b\mu \nu } \partial _\mu   h _{\nu \beta \gamma
}^c  +
\partial _\beta  \varepsilon ^{b\mu \nu } \partial _\mu   h
_{\alpha \nu \gamma }^c  + \partial _\gamma  \varepsilon ^{b\mu \nu
} \partial _\mu   h _{\alpha \beta \nu }^c
\nonumber\\
 && + \partial _\alpha  \partial _\beta  \varepsilon ^{b\mu \nu }  h _{\mu \nu \gamma }^c  + \partial _\alpha
 \partial _\gamma  \varepsilon ^{b\mu \nu }  h _{\mu \beta \nu }^c  + \partial _\beta  \partial _\gamma  \varepsilon ^{b\mu \nu }  h _{\alpha \mu \nu }^c )\label{3.7}
\end{eqnarray}
This form of $\delta _\varepsilon ^1  h _{\alpha \beta \gamma }^a$ is not unique at this stage of Noether's procedure and defined due to $\delta _\varepsilon ^0  h _{\alpha \beta \gamma }^a$ with linearly field dependent gauge parameter. This can be easily seen comparing (\ref{3.4}) and (\ref{3.3}).

Inserting (\ref{3.6}) and (\ref{3.7}) into Noether's equation (\ref{3.4}) we obtain the following nice result
\begin{eqnarray}
&&\delta^{0}_\varepsilon
R^{a(2)}_{\alpha\alpha',\beta\beta',\gamma\gamma'}+\delta^{1}_\varepsilon
R^{a(1)}_{\alpha\alpha',\beta\beta',\gamma\gamma'}=f^{abc}\big(
\varepsilon^{b\mu\nu}\partial_{\mu}\partial_{\nu}R^{c(1)}_{\alpha\alpha',\beta\beta',\gamma\gamma'}\nonumber\\&&+
\partial_{\alpha}\varepsilon^{b\mu\nu}\partial_{\nu}R^{c(1)}_{\mu\alpha',\beta\beta',\gamma\gamma'}+
\partial_{\alpha'}\varepsilon^{b\mu\nu}\partial_{\nu}R^{c(1)}_{\alpha\mu,\beta\beta',\gamma\gamma'}+
\partial_{\beta}\varepsilon^{b\mu\nu}\partial_{\nu}R^{c(1)}_{\alpha\alpha',\mu\beta',\gamma\gamma'}\nonumber\\&&+
\partial_{\beta'}\varepsilon^{b\mu\nu}\partial_{\nu}R^{c(1)}_{\alpha\alpha',\beta\mu,\gamma\gamma'}+
\partial_{\gamma}\varepsilon^{b\mu\nu}\partial_{\nu}R^{c(1)}_{\alpha\alpha',\beta\beta',\mu\gamma'}+
\partial_{\gamma'}\varepsilon^{b\mu\nu}\partial_{\nu}R^{c(1)}_{\alpha\alpha',\beta\beta',\gamma\mu}\nonumber\\&&+
\partial_{\alpha}\partial_{\beta}\varepsilon^{b\mu\nu}R^{c(1)}_{\mu\alpha',\nu\beta',\gamma\gamma'}+
\partial_{\alpha}\partial_{\gamma}\varepsilon^{b\mu\nu}R^{c(1)}_{\mu\alpha',\beta\beta',\nu\gamma'}+
\partial_{\beta}\partial_{\gamma}\varepsilon^{b\mu\nu}R^{c(1)}_{\alpha\alpha',\mu\beta',\nu\gamma'}\nonumber\\&&+
\partial_{\alpha'}\partial_{\beta'}\varepsilon^{b\mu\nu}R^{c(1)}_{\alpha\mu,\beta\nu,\gamma\gamma'}+
\partial_{\alpha'}\partial_{\gamma'}\varepsilon^{b\mu\nu}R^{c(1)}_{\alpha\mu,\beta\beta',\gamma\nu}+
\partial_{\beta'}\partial_{\gamma'}\varepsilon^{b\mu\nu}R^{c(1)}_{\alpha\alpha',\beta\mu,\gamma\nu}\nonumber\\&&+
\partial_{\alpha'}\partial_{\beta}\varepsilon^{b\mu\nu}R^{c(1)}_{\alpha\mu,\nu\beta',\gamma\gamma'}+
\partial_{\alpha'}\partial_{\gamma}\varepsilon^{b\mu\nu}R^{c(1)}_{\alpha\mu,\beta\beta',\nu\gamma'}+
\partial_{\beta'}\partial_{\gamma}\varepsilon^{b\mu\nu}R^{c(1)}_{\alpha\alpha',\beta\mu,\nu\gamma'}\nonumber\\&&+
\partial_{\beta'}\partial_{\alpha}\varepsilon^{b\mu\nu}R^{c(1)}_{\nu\alpha',\beta\mu,\gamma\gamma'}+
\partial_{\gamma'}\partial_{\alpha}\varepsilon^{b\mu\nu}R^{c(1)}_{\mu\alpha',\beta\beta',\gamma\nu}+
\partial_{\gamma'}\partial_{\beta}\varepsilon^{b\mu\nu}R^{c(1)}_{\alpha\alpha',\mu\beta',\gamma\nu}
\big)\label{3.8}
\end{eqnarray}
So we see that interpreting the right hand side of (\ref{3.7}) as a generalization of the Lie derivative of symmetric covariant tensors, we obtain the r.h.s. of (\ref{3.8}) as constructed in the same way as "Lie derivative" of the first order curvature $R^{a(1)}_{\alpha\alpha',\beta\beta',\gamma\gamma'}$ with three pairs of antisymmetrized indices.
Finally note also that the first order transformation (\ref{3.7}) can be rewritten in the following form
\begin{eqnarray}
\delta _\varepsilon ^1  h _{\alpha \beta \gamma }^a=&&f^{abc}
\big[\partial _\alpha(
\varepsilon ^{b\mu \nu } \partial _\mu   h _{\nu \beta \gamma
}^c -  \varepsilon ^{b\mu \nu } \partial _\beta h _{\mu \nu \gamma }^c - \varepsilon ^{b\mu \nu }\partial _\gamma  h _{\mu \nu \beta }^c)\nonumber\\
 &&\,\,\quad +\partial _\beta  (\varepsilon ^{b\mu \nu } \partial _\mu   h
_{\nu\gamma\alpha }^c - \varepsilon ^{b\mu \nu } \partial _\gamma  h _{\mu \nu \alpha }^c -   \varepsilon ^{b\mu \nu }  \partial _\alpha h _{\mu \nu \gamma }^c)\nonumber\\&&\,\,\quad  + \partial _\gamma  (\varepsilon ^{b\mu \nu
} \partial _\mu   h _{\nu\alpha \beta }^c  -
 \varepsilon ^{b\mu \nu } \partial _\alpha h _{\mu \nu \beta }^c -
 \varepsilon ^{b\mu \nu } \partial _\beta h _{\mu \nu \alpha }^c )\nonumber\\
 &&+\partial _\alpha \partial _\beta (\varepsilon ^{b\mu \nu } h _{\mu \nu \gamma }^c) + \partial _\beta \partial _\gamma (\varepsilon ^{b\mu \nu } h _{\mu \nu \alpha }^c) + \partial _\gamma  \partial _\alpha (\varepsilon ^{b\mu \nu } h _{\mu \nu \beta }^c )\nonumber\\
 && \,\,\quad + \varepsilon ^{b\mu \nu } \Gamma^{c}_{\mu\nu,\alpha\beta\gamma}\big]\label{3.9}
\end{eqnarray}
and therefore we can separate in (\ref{3.9}) at this stage inessential symmetrized gradients (i.e. $\delta^0$ with field dependent parameter) and obtain the essential part of $\delta _\varepsilon ^1  h _{\alpha \beta \gamma }^a$ in the following elegant form:
\begin{equation}\label{3.10}
    \tilde{\delta} _\varepsilon ^1  h _{\alpha \beta \gamma }^a = f^{abc}\varepsilon ^{b\mu \nu } \Gamma^{c}_{\mu\nu,\alpha\beta\gamma}
\end{equation}

\section{The general spin $s$ case}
\setcounter{equation}{0}
\quad To start the work with general symmetric tensors we follow the notations of our previous papers ( see \cite{Manvelyan:2008ks} and ref. there) and introduce an additional formal vector variable $a^{\mu}$ to handle rank $s$ symmetric tensors as the monomials
\begin{equation}\label{4.1}
   h^{(s)}(x;a)= h_{\mu_{1}\mu_{2}\dots\mu_{s}}(x) a^{\mu_{1}}a^{\mu_{2}}\dots a^{\mu_{s}}
\end{equation}
In these notes we need to define only two operations to perform all calculations:
\begin{itemize}
  \item Symmetrized gradient
  \begin{equation}\label{4.2}
    \partial_{(\mu_{s+1}}h_{\mu_{1}\mu_{2}\dots\mu_{s})}\Rightarrow (a\nabla) h^{(s)}(x;a)
  \end{equation}
  \item Contraction inside the set of symmetrized indices (star product):
  \begin{eqnarray}
    &&T(x)_{\mu_{1}\mu_{2}\dots\mu_{s}}H^{\mu_{1}\mu_{2}\dots\mu_{s}}(x) \Rightarrow T^{(s)}(x;a)*_{a}H^{(s)}(x;a). \nonumber\\
    && \textnormal{where}\quad *_{a}=\frac{1}{(s!)^{2}} \prod^{s}_{i=1}\overleftarrow{\partial}^{\mu_{i}}_{a}\overrightarrow{\partial}_{\mu_{i}}^{a} .\label{4.3}
  \end{eqnarray}
\end{itemize}
To distinguish easily between "a" and "x" spaces we introduce for space-time derivatives $\frac{\partial}{\partial x^{\mu}}$ the notation $\nabla_{\mu}$, $(a\nabla)=a^{\mu}\nabla_{\mu}$.

Then using these notations we can write a zero order gauge transformation (symmetrized gradient) in the following form:
\begin{equation}\label{4.4}
    \delta_{(0)}h^{(s)}(x;a)=(a\nabla)\epsilon^{(s-1)}(x;a)
\end{equation}
where $\epsilon^{(s-1)}(x;a)$ is a rank $s-1$ symmetric tensor gauge parameter for the spin $s$ gauge field\footnote{We do not necessarily discuss constrained higher spin fields here (Fronsdal's formulation \cite{Frons}), our discussion is relevant also for unconstrained higher spin fields representing general rank $s$ symmetric tensors.}. Other important expressions are the hierarchy of first order generalized Christoffel symbols introduced in \cite{DF}. These $(n,s)$ bitensors
\begin{equation}\label{4.5}
\Gamma^{(n)}_{(1)}(x;b,a)\equiv \Gamma^{(n)}_{(1)}(x)_{\rho_{1}...\rho_{n},\mu_{1}...\mu_{s}}b^{\rho_{1}}...b^{\rho_{n}}a^{\mu_{1}}...a^{\mu_{s}}
\end{equation}
can be written in our notation in an elegant form:
\begin{equation}\label{4.6}
 \Gamma^{(n)}_{(1)}(x;b,a)=\sum_{k=0}^{n}\frac{(-1)^{k}}{k!}(b\nabla)^{n-k}(a\nabla)^{k}(b\partial_{a})^{k}h^{(s)}(x;a)
\end{equation}
Inserting (\ref{4.4}) into the latter we obtain the transformation law for these objects:
\begin{eqnarray}
&&\delta_{(0)} \Gamma^{(n)}_{(1)}(x;b,a)=\frac{(-1)^{n}}{n!}(a\nabla)^{n+1}(b\partial_{a})^{n}\epsilon^{(s-1)}(x;a) ,\label{4.7}
\end{eqnarray}
So we see that the last term of the hierarchy of the linearized curvature
\begin{equation}\label{4.8}
  R_{(1)}(x;b,a)=\Gamma^{(s)}_{(1)}(x;b,a)
  =\sum_{k=0}^{s}\frac{(-1)^{k}}{k!}(b\nabla)^{s-k}(a\nabla)^{k}(b\partial_{a})^{k}h^{(s)}(x;a)
\end{equation}
is invariant with respect to the gauge transformation (\ref{4.4}):
\begin{equation}\label{4.9}
    \delta_{(0)} R_{(1)}(x;b,a)=0
\end{equation}

Another important object for the present considerations is the last one before the curvature Christoffel symbol $\Gamma(x;b,a)=\Gamma^{s-1}_{(1)}(x;b,a)$ with the gauge transformation
\begin{equation}\label{4.10}
    \delta_{(0)} \Gamma(x;b,a)=(-1)^{s-1}(a\nabla)^{s}\epsilon^{(s-1)}(x;b)
\end{equation}
It is worth to note that $\Gamma(x;b,a)$ reproduces correctly (\ref{2.7}) for $s=2$ and (\ref{3.5}) for $s=3$.
The expression (\ref{4.8}) can be written in another useful form:
\begin{equation}\label{4.11}
    R_{(1)}(x;b,a)=\frac{1}{(s!)^{2}}\left[(b\partial_{d})(a\partial_{e})-(b\partial_{e})(a\partial_{d})\right]^{s} (d\nabla)^{s}h^{(s)}(x;e)
\end{equation}
Finally using (\ref{4.6}), (\ref{4.8}) and (\ref{4.11}) we present the connection between $\Gamma(x;b,a)$ and curvature $R_{(1)}(x;b,a)$
\begin{equation}\label{4.12}
    \left[(b\nabla)(a\partial_{e})-(a\nabla)(b\partial_{e})\right]\Gamma(x;c,e)=\frac{1}{s+1} \left[(b\partial_{c})(a\partial_{e})-(a\partial_{c})(b\partial_{e})\right]R_{(1)}(x;c,e)
\end{equation}
Now we are ready to derive the second order curvature. First we propose the following generalization of (\ref{3.10}) and (\ref{2.14}) for the essential part of the first order spin $s$ gauge transformation:
\begin{equation}\label{4.13}
    \delta_{(1)}h^{(s)}(x;a)=\epsilon^{(s-1)}(x;c)*_{c}\Gamma(x;c,a)
\end{equation}
Then we calculate the first order variation of the first order curvature (\ref{4.11})
\begin{eqnarray}
  \delta_{(1)}R_{(1)}(x;b,a)&=&\frac{1}{(s!)^{2}}
  \left[(b\partial_{d})(a\partial_{e})-(b\partial_{e})(a\partial_{d})\right]^{s} \left\{\left[(d\nabla)^{s}\epsilon^{(s-1)}(x;c)\right]*_{c}\Gamma(x;c,e)\right. \nonumber\\
  &+&\left. \sum^{s}_{k=1}\binom{s}{k} \left[(d\nabla)^{s-k}\epsilon^{(s-1)}(x;c)\right]*_{c}(d\nabla)^{k}\Gamma(x;c,e)\right\}\label{4.14}
\end{eqnarray}
Applying (\ref{4.10}) and (\ref{4.12}) we can rewrite the latter variation in the following way:
\begin{eqnarray}
  \delta_{(1)}R_{(1)}(x;b,a)&&=\frac{(-1)^{s-1}}{(s!)^{2}}A(b,a;\partial_{d},\partial_{e})^{s} \delta_{(0)} \Gamma(x;c,d)*_{c}\Gamma(x;c,e) \nonumber\\
  &&+\frac{1}{(s+1)!}\sum^{s}_{k=1}\binom{s}{k} [A(b,a;\nabla,\partial_{e})]^{s-k}\epsilon^{(s-1)}(x;c)*_{c}\nonumber\\&&[A(b,a;\nabla,\partial_{e})]^{k-1}
  A(b,a;\partial_{c},\partial_{e})R_{(1)}(x;c,e)\label{4.15}
\end{eqnarray}
where
\begin{eqnarray}
   A(b,a;\nabla,\partial_{e})&=&(b\nabla)(a\partial_{e})-(a\nabla)(b\partial_{e}) \label{4.16}\\
   A(b,a;\partial_{c},\partial_{e})&=&(b\partial_{c})(a\partial_{e})-(a\partial_{c})(b\partial_{e}) \label{4.17}
\end{eqnarray}
So looking at (\ref{4.15}) we see that after functional integration in the first line we obtain a solution of the expected covariant Noether's equation:
\begin{eqnarray}
  && \delta_{(1)} R_{(1)}(x;b,a)+\delta_{(0)} R_{(2)}(x;b,a)=\mathcal{L}_{\epsilon}R_{(1)}(x;b,a)\label{4.18}
\end{eqnarray}
where
\begin{eqnarray}
  R_{(2)}(x;b,a)&=&\frac{1}{2(s!)^{2}}\left[(b\partial_{d})(a\partial_{e})
                                                        -(b\partial_{e})(a\partial_{d})\right]^{s}
\Gamma(x;c,d)*_{c}\Gamma(x;c,e)\qquad\label{4.19}
\end{eqnarray}
is the second order curvature and
\begin{eqnarray}
  \mathcal{L}_{\epsilon}R_{(1)}(x;b,a)
   &=&\frac{1}{(s+1)!}\sum^{s}_{k=1}\binom{s}{k}  [A(b,a;\nabla,\partial_{e})]^{s-k}\epsilon^{(s-1)}(x;c)\nonumber\\
   && *_{c}[A(b,a;\nabla,\partial_{e})]^{k-1}A(b,a;\partial_{c},\partial_{e})R_{(1)}(x;c,e)\label{4.20}
\end{eqnarray}
is the generalized Lie derivative or spin $s$ reparametrization. The formulas (\ref{4.12}) and (\ref{4.19}) make (\ref{1.1}) obvious. It is also obvious from (\ref{4.19}) that odd spin fields require additional Yang-Mills like internal symmetry indices for nontriviality.

To compare the results of this paper to the ones obtained in AdS space by Vasiliev and collaborators (see \cite{VasilievEqn} and references therein), one should continue this results from flat space to the AdS as it was done for linear curvatures in \cite{MR5, MR6}.
The AdS continuation of the obtained second order curvature is relevant also to AdS/CFT tasks (see \cite{Manvelyan:2008ks}, \cite{Kleb}, \cite{Ruehl}, \cite{Giombi:2010vg}), and we hope to be able to do it in near future.

\subsection*{Acknowledgements}

K.M. is grateful to E. Skvortsov for useful comments.
This work is supported in part by Alexander von Humboldt Foundation under 3.4-Fokoop-ARM/1059429.
Work of K.M. was made with partial support of CRDF-NFSAT-SCS MES RA ECSP 09\_01/A-31.

\end{document}